\documentclass[intlimits,twoside,a4paper]{article}
\usepackage{amsmath,amssymb}
\usepackage{graphicx}
\usepackage{wrapfig}
\usepackage[T2A]{fontenc}
\usepackage[cp1251]{inputenc}

\usepackage{color}

\usepackage{cmpj2}

\DeclareMathOperator{\sgn}{sgn}
\DeclareMathOperator{\Ai}{Ai}

\issue{2014}{17}{3}{33003}
\doinumber{10.5488/CMP.17.33003}

\title[$N$-point free energy distribution function  in one dimensional random directed polymers]%
{$N$-point free energy distribution function \\ in one dimensional random directed polymers}

\author[V. Dotsenko]{V. Dotsenko\refaddr{label1,label2}}
\addresses{
\addr{label1} LPTMC, Universit\'e Paris VI, 75252 Paris, France \\
\addr{label2} L.D. Landau Institute for Theoretical Physics, 119334 Moscow, Russia
}

\date{Received April 16, 2014, in final form May 14, 2014}
\authorcopyright{V. Dotsenko, 2014}

\begin{document}

\maketitle

\begin{abstract}

Explicit expression for the $N$-point free energy distribution function in one dimensional
directed polymers in a random potential is derived in terms of the Bethe ansatz replica
technique. The obtained result is equivalent to the one
derived earlier by Prolhac and Spohn  [J. Stat. Mech., 2011, P03020].



\end{abstract}

\section{Introduction}

In this paper we consider the model of one-dimensional directed polymers in a quenched random potential.
This model is defined in terms of an elastic string $\phi(\tau)$
directed along the $\tau$-axis within an interval $[0,t]$ which passes through a random medium
described by a random potential $V(\phi,\tau)$. The energy of a given polymer's trajectory
$\phi(\tau)$ is

\begin{equation}
   \label{1}
   H[\phi(\tau), V] = \int_{0}^{t} \rd\tau
   \left\{\frac{1}{2} \bigl[\partial_\tau \phi(\tau)\bigr]^2
   + V[\phi(\tau),\tau]\right\},
\end{equation}
where the disorder potential $V[\phi,\tau]$ is described by the Gaussian distribution
with a zero mean $\overline{V(\phi,\tau)}=0$
and the $\delta$-correlations: ${\overline{V(\phi,\tau)V(\phi',\tau')}} = u \delta(\tau-\tau') \delta(\phi-\phi')$
The parameter $u$ describes the strength of the disorder.

The system of this type as well as the equivalent problem of the KPZ-equation
\cite{KPZ} describing the growth of an interface with time
in the presence of noise
have been the subject of intense investigations for about the last three
decades (see e.g.
\cite{hh_zhang_95,burgers_74,kardar_book,hhf_85,numer1,numer2,kardar_87,bouchaud-orland,Brunet-Derrida,
Johansson,Prahofer-Spohn,Ferrari-Spohn1}).
Such a system exhibits numerous non-trivial features due to the interplay
between elasticity and disorder. In particular, in the  limit $t \to \infty$,
the polymer mean squared displacement exhibits a universal scaling form
$\overline{\langle\phi^{2}\rangle} \propto t^{4/3} $
(where $\langle \dots \rangle$ and $\overline{(\dots)}$ denote
the thermal and the disorder averages) while
the typical value of the free energy fluctuations scales as $t^{1/3}$.
Note that in the corresponding pure system (with $V(\phi,\tau) \equiv 0$)
$\langle\phi^{2}\rangle \propto t$ while the free energy is proportional to $\ln(t)$.

A few years ago, an exact solution for the free energy
probability distribution function (PDF)  has been found
\cite{KPZ-TW1a,KPZ-TW1b,KPZ-TW1c,KPZ-TW2,BA-TW1,BA-TW2,BA-TW3,LeDoussal1,LeDoussal2,LeDoussal2',goe,LeDoussal3,Corwin,Borodin}.
It was shown that depending on the boundary conditions, this PDF is given by the Tracy-Widom (TW) distribution
\cite{TW-GUE} either of the Gaussian Unitary Ensemble (GUE) or of the Gaussian Orthogonal Ensemble (GOE) or
of the Gaussian Simplectic Ensemble (GSE).
Besides, recently the two-point free energy distribution function
which describes the joint statistics of the free energies of the directed polymers
coming to two different endpoints has been derived in  \cite{Prolhac-Spohn,2pointPDF,Imamura-Sasamoto-Spohn}.

For fixed boundary conditions, $\phi(0) = 0; \; \phi(t) = x$, the partition function
of the model (\ref{1}) is
\begin{equation}
\label{2}
   Z_{t}(x) = \int_{\phi(0)=0}^{\phi(t)=x}
              {\cal D} \phi(\tau)  \;  \re^{-\beta H[\phi]}
\; = \; \exp\left[-\beta F_{t}(x)\right] \, ,
\end{equation}
where $\beta$ is the inverse temperature and $F_{t}(x)$ is the free energy.
In the limit $t\to\infty$, the free energy scales as
\begin{equation}
\label{3}
 \beta F_{t}(x) = \beta f_{0} t + \beta x^{2}/2t + \lambda_{t} f(x) \, ,
\end{equation}
where $f_{0}$ is the selfaveraging free energy density and
\begin{equation}
 \label{4}
\lambda_{t} = \frac{1}{2} \left(\beta^{5}u^{2} t\right)^{1/3} \propto t^{1/3} \, .
\end{equation}
It is the statistics of rescaled free energy fluctuations $f(x)$
which in the limit $t\to\infty$ is
expected to be described by a non-trivial universal distribution $W(f)$.
In fact, the first two trivial terms of this free energy can be easily
eliminated by simple redefinition of the partition function:
\begin{equation}
 \label{5}
 Z_{t}(x) \to \exp\left\{-\beta f_{0} t - \beta x^{2}/2t \right\} \, \tilde{Z}_{t}(x)
\end{equation}
so that
\begin{equation}
 \label{6}
 \tilde{Z}_{t}(x) \; = \; \exp\bigl\{-\lambda_{t} f(x)\bigr\} \, .
\end{equation}

The aim of the present work is to  study the $N$-point free energy
probability distribution function
\begin{equation}
\label{7}
W(f_{1}, \ldots , f_{N}; x_{1}, \ldots , x_{N})\; \equiv \; W({\bf f}; {\bf x}) \; = \;
\lim_{t\to\infty} \; \mathrm{Prob}\bigl[f(x_{1})>f_{1}, \, \ldots \, , f(x_{N})>f_{N}\bigr] \, ,
\end{equation}
which describes the joint statistics of the free energies of $N$ directed polymers
coming to $N$ different endpoints. Some time ago the result for this function
has been derived in terms of the Bethe ansatz replica technique under a particular
decoupling assumption \cite{Prolhac-Spohn-N}. Here, I am going to recompute this
function using somewhat different computational tricks which do not require
any supplementary assumptions and which permit to represent the final result in somewhat
more explicit form.

\section{$N$-point distribution function}

The probability distribution function,
equation (\ref{7}) can be defined as follows:
\begin{equation}
\label{8}
W({\bf f}; {\bf x}) = \lim_{\lambda\to\infty}
\sum_{L_{1},\ldots,L_{N}=0}^{\infty}
\prod_{k=1}^{N}\Biggl[
\frac{(-1)^{L_{k}}}{L_{k}!}
\exp\bigl(\lambda L_{k} f_{k} \bigr) \Biggr]\;
\overline{\Biggl(\prod_{k=1}^{N} \tilde{Z}_{t}(x_{k})\Biggr)} \, ,
\end{equation}
where $\overline{(\ldots)}$ denotes the average over random
potentials. Indeed, substituting here equation (\ref{6}) we get
\begin{equation}
 \label{9}
 W({\bf f}; {\bf x}) = \lim_{\lambda\to\infty}
 \overline{\Biggl(
 \prod_{k=1}^{N}
 \exp\Bigl\{-\exp\Bigl[\lambda_{t} \bigl(f_{k} - f(x_{k})\bigr)\Bigr]\Bigr\} \; \Biggr) } \; = \;
 \overline{\Biggl[
 \prod_{k=1}^{N} \theta\bigl(f(x_{k}) - f_{k} \bigr) \Biggr]}
\end{equation}
which coincides with the definition (\ref{7}).

Performing the standard  averaging over random potentials in equation (\ref{8})
one obtains (for details see e.g. \cite{BA-TW3})
\begin{equation}
\label{10}
W({\bf f}; {\bf x}) = \lim_{\lambda\to\infty}
\sum_{L_{1},\ldots,L_{N}=0}^{\infty}
\prod_{k=1}^{N}\Biggl[
\frac{(-1)^{L_{k}}}{L_{k}!}
\exp\bigl(\lambda L_{k} f_{k} \bigr) \Biggr]\;
\Psi\bigl(\underbrace{x_{1}, \ldots,x_{1}}_{L_{1}}, \,
          \underbrace{x_{2}, \ldots,x_{2}}_{L_{2}}, \, \ldots\, ,
          \underbrace{x_{N}, \ldots,x_{N}}_{L_{N}} ; \; t\bigr)\,,
\end{equation}
where the time dependent $n$-point wave function $\Psi(x_{1}, \ldots, x_{n}; t)$
($n = \sum_{k=1}^{N} L_{k}$)
is the solution of the
imaginary time Schr\"odinger equation
\begin{equation}
 \label{11}
\beta \, \partial_t \Psi({\bf x}; t) \; = \;
\left[\frac{1}{2}\sum_{a=1}^{n}\partial_{x_a}^2
   +\frac{1}{2}\, \kappa \sum_{a\not=b}^{n} \delta(x_a-x_b)\right]
\Psi({\bf x}; t)
\end{equation}
with $\kappa = \beta^{3}u$ and  the initial condition
\begin{equation}
   \label{12}
\Psi({\bf x}; t=0) = \prod_{a=1}^{n} \delta(x_a) \, .
\end{equation}
A generic  eigenstate of such a system is characterized by $n$ momenta
$\{ Q_{a} \} \; (a=1,\ldots,n)$ which split into
$M$  ($1 \leqslant M \leqslant n$) clusters described by
continuous real momenta $q_{\alpha}$ $(\alpha = 1,\ldots,M)$
and having $n_{\alpha}$ discrete imaginary parts
\begin{equation}
   \label{13}
Q_{a} \; \equiv \; q^{\alpha}_{r} \; = \;
q_{\alpha} - \frac{\ri\kappa}{2}  (n_{\alpha} + 1 - 2r),
\qquad
(r = 1, \ldots, n_{\alpha}),
\end{equation}
with the global constraint
\begin{equation}
   \label{14}
\sum_{\alpha=1}^{M} n_{\alpha} = n \, .
\end{equation}
The time dependent solution  $\Psi({\bf x},t)$
of the Schr\"odinger equation (\ref{11}) with the initial conditions, equation (\ref{12}),
can be represented in the form of a linear combination of eigenfunctions
$\Psi_{\bf Q}^{(M)}({\bf x})$:
\begin{equation}
   \label{15}
\Psi({\bf x}; t) =
\sum_{M=1}^{N} \frac{1}{M!}
\prod_{\alpha=1}^{M} \left[\int_{-\infty}^{+\infty} \frac{\rd q_{\alpha}}{2\pi} \sum_{n_{\alpha}=1}^{\infty}\right]
{\boldsymbol \delta}\left(\sum_{\alpha=1}^{M} n_{\alpha}, \, n\right) \,
\frac{\kappa^{N}|C_{M}({\bf q},{\bf n})|^{2}}{N! \prod_{\alpha=1}^{M}\bigl(\kappa n_{\alpha}\bigr)} \,
\Psi^{(M)}_{{\bf Q}}({\bf x})
{\Psi^{(M)}_{{\bf Q}}}^{*}({\bf 0}) \;
\exp\bigl\{-E_{M}({\bf q},{\bf n}) t \bigr\} \, .
\end{equation}
Here, ${\boldsymbol \delta}(k , m)$ is the Kronecker symbol,
the normalization factor
\begin{equation}
   \label{16}
|C_{M}({\bf q}, {\bf n})|^{2} =
\prod_{\alpha<\beta}^{M}
\frac{\big|q_{\alpha}-q_{\beta} -\frac{\ri\kappa}{2}(n_{\alpha}-n_{\beta})\big|^{2}}{
      \big|q_{\alpha}-q_{\beta} -\frac{\ri\kappa}{2}(n_{\alpha}+n_{\beta})\big|^{2}}
\end{equation}
and the eigenvalues:
\begin{equation}
\label{17}
E_{M}({\bf q},{\bf n}) \; = \;
\sum_{\alpha=1}^{M}
\left(
\frac{1}{2\beta} n_{\alpha} q_{\alpha}^{2}
- \frac{\kappa^{2}}{24\beta} n_{\alpha}^{3}
\right) \, .
\end{equation}
For a given set of integers $\{M; n_{1}, \ldots., n_{M}\}$,
the eigenfunctions $\Psi_{\bf Q}^{(M)}({\bf x})$ can be represented as follows
(for details see \cite{Lieb-Liniger,McGuire,Yang,Calabrese,rev-TW}):
\begin{equation}
\label{18}
\Psi^{(M)}_{{\bf q}}({\bf x}) =
\sum_{{\cal P}}  \;
\prod_{a<b}^{n}
\left[
1 +\ri \kappa \frac{\sgn(x_{a}-x_{b})}{Q_{{\cal P}_a} - Q_{{\cal P}_b}}
\right] \;
\exp\left(\ri \sum_{a=1}^{n} Q_{{\cal P}_{a}} x_{a} \right)\,,
\end{equation}
where the summation goes over $n!$ permutations ${\cal P}$ of $n$ momenta $Q_{a}$,
equation (\ref{13}),  over $n$ particles $x_{a}$.

Substituting equations (\ref{15})--(\ref{18}) into equation (\ref{10}) we get
\begin{align}
 \label{19}
 W({\bf f}; {\bf x}) =& 1 + \lim_{\lambda\to\infty} \Biggl\{
\sum_{L_{1}+\ldots+L_{N}\geqslant 1}^{\infty}
\prod_{k=1}^{N}\left[
\frac{(-1)^{L_{k}}}{L_{k}!}
\exp\bigl(\lambda L_{k} f_{k} \bigr) \right]\;
\nonumber
\\[1ex]
\nonumber
&\times
\sum_{M=1}^{L_{1}+\ldots+L_{N}} \frac{1}{M!}
\prod_{\alpha=1}^{M}
\left[
\sum_{n_{\alpha}=1}^{\infty}
\int_{-\infty}^{+\infty} \rd q_{\alpha} \frac{\kappa n_{\alpha}}{2\pi \kappa n_{\alpha}}
\exp\left(
-\frac{t}{2\beta} n_{\alpha} q_{\alpha}^{2}
+ \frac{\kappa^{2} t}{24 \beta} n_{\alpha}^{3}
\right)
\right]
 {\boldsymbol \delta}\left(\sum_{\alpha=1}^{M} n_{\alpha}  ,  \sum_{k=1}^{N}L_{k}\right)
 |C_{M}({\bf q}, {\bf n})|^{2}
\\[1ex]
&\times
\sum_{{\cal P}^{(L_{1},\ldots,L_{N})}}
\prod_{k=1}^{N}
\left[
\sum_{{\cal P}^{(L_{k})}}
\right]
\prod_{k<l}^{N} \prod_{a_{k}=1}^{L_{k}} \prod_{a_{l}=1}^{L_{l}}
\left(
\frac{Q_{{\cal P}_{a_{k}}^{(L_{k})}} - Q_{{\cal P}_{a_{l}}^{(L_{l})}} - \ri \kappa}{
      Q_{{\cal P}_{a_{k}}^{(L_{k})}} - Q_{{\cal P}_{a_{l}}^{(L_{l})}}}
\right)
\exp\left(\ri \sum_{k=1}^{N} x_{k} \sum_{a_{k}=1}^{L_{k}} Q_{{\cal P}_{a_{k}}^{(L_{k})}}\right)
\Biggr\} \, .
\end{align}

In the above expression, the summation over permutations of $n = L_{1} + \ldots + L_{N}$ momenta $Q_{a}$
split into the internal permutations ${\cal P}^{(L_{k})}$ of $L_{k}$ momenta
[taken at random out of the total list $\{ Q_{a} \} \; (a=1, \ldots, n)$] and the permutations
${\cal P}^{(L_{1},\ldots,L_{N})}$ of the momenta among the groups $L_{k}$.
It is evident that due to the symmetry of the expression in equation (\ref{19}), the summations over
${\cal P}^{(L_{k})}$ give just the factor $L_{1}! \ldots L_{N}!$.
On the other hand, the structure of the
Bethe ansatz wave functions, equation (\ref{18}), is such that
for the positions of ordered particles
in the summation over permutations, the momenta $Q_{a}$ belonging
to the same cluster also remain ordered (for details see e.g. \cite{rev-TW}).
Thus, in order to perform the summation over the permutations ${\cal P}^{(L_{1},\ldots,L_{N})}$
in equation (\ref{19}) it is sufficient to split the momenta of each cluster into $N$ parts:
\begin{equation}
 \label{20}
 \{
\underbrace{q_{1}^{\alpha}, \ldots,  q_{m^{1}_{\alpha}}^{\alpha}}_{m^{1}_{\alpha}} ; \;
\underbrace{q_{m^{1}_{\alpha}+1}^{\alpha}, \ldots,  q_{m^{1}_{\alpha}+m^{2}_{\alpha}}^{\alpha}}_{m^{2}_{\alpha}}  ; \; \ldots \; ; \;
\underbrace{q_{\sum_{k=1}^{N-1}m^{k}_{\alpha}+1}^{\alpha}, \ldots,  q_{\sum_{k=1}^{N}m^{k}_{\alpha}}^{\alpha}}_{m^{N}_{\alpha}}  \}\, ,
\end{equation}
where the integers $m^{k}_{\alpha} = 0, 1, \ldots , n_{\alpha}$ are constrained by the conditions
\begin{eqnarray}
 \label{21}
 \sum_{k=1}^{N} m^{k}_{\alpha} &=& n_{\alpha}\,,
 \\
 \label{22}
 \sum_{\alpha=1}^{M} m^{k}_{\alpha} &=& L_{k}\,,
\end{eqnarray}
and the momenta of every group
$\left\{ q_{\sum_{l=1}^{k-1}m^{l}_{\alpha}+1}^{\alpha}, \; \ldots, \;  q_{\sum_{l=1}^{k}m^{l}_{\alpha}}^{\alpha} \right\}$
all belong to the particles whose coordinates are all equal to $x_{k}$.
Let us redefine:
\begin{equation}
 \label{23}
 q_{\sum_{l=1}^{k-1}m^{l}_{\alpha}+r}^{\alpha} \; \equiv \;
 q_{k,r}^{\alpha} \; = \; q_{\alpha} + \frac{\ri\kappa}{2}\left(n_{\alpha} + 1 - 2 \sum_{l=1}^{k-1}m^{l}_{\alpha} -2r\right).
\end{equation}
In this way, the summation over ${\cal P}^{(L_{1},\ldots,L_{N})}$ is changed by the summation over
the integers $\{m^{k}_{\alpha}\}$. Substituting equations (\ref{20})--(\ref{23}) into equation (\ref{19}) after simple
algebra, we find
\begin{align}
 \label{24}
W({\bf f}; {\bf x}) =& 1 + \lim_{\lambda\to\infty} \left(
\sum_{M=1}^{\infty} \frac{(-1)^{M}}{M!}
\prod_{\alpha=1}^{M}
\left\{
\sum_{\sum_{k}^{N}m^{k}_{\alpha} \geqslant 1} (-1)^{\sum_{k}^{N}m^{k}_{\alpha} -1}
\int_{-\infty}^{+\infty} \frac{\rd q_{\alpha}}{2\pi\kappa \bigl(\sum_{k}^{N}m^{k}_{\alpha} \bigr)} \right.\right.
\nonumber
\\[1ex]
\nonumber
&\times
\left.\exp\left[\lambda\sum_{k=1}^{N}m^{k}_{\alpha} f_{k} +\ri \sum_{k=1}^{N}m^{k}_{\alpha} x_{k} q_{\alpha}
-\frac{1}{4} \kappa \sum_{k,l=1}^{N}m^{k}_{\alpha}m^{l}_{\alpha} \big|x_{k} - x_{l}\big|
-\frac{t}{2\beta} q_{\alpha}^{2} \sum_{k=1}^{N}m^{k}_{\alpha}
+\frac{\kappa^{2} t}{24\beta} \left(\sum_{k=1}^{N}m^{k}_{\alpha}\right)^{3} \right]
\right\}
\\[1ex]
&\times
\big|C_{M}\bigl({\bf q}; \{ m^{k}_{\alpha}\}\bigr)\big|^{2} \; G_{M}\bigl({\bf q}; \{ m^{k}_{\alpha}\}\bigr)\left.\vphantom{\int_{-\infty}^{+\infty}}\right)\,,
\end{align}
where the normalization constant $\big|C_{M}\bigl({\bf q}; \{ m^{k}_{\alpha}\}\bigr)\big|^{2}$
is given in equation (\ref{16}) (with $n_{\alpha} = \sum_{k=1}^{N}m^{k}_{\alpha}$) and
\begin{equation}
 \label{25}
 G_{M}\bigl({\bf q}; \{ m^{k}_{\alpha}\}\bigr) =
 \prod_{\alpha=1}^{M} \prod_{k<l}^{N}\prod_{r=1}^{m^{k}_{\alpha}} \prod_{r'=1}^{m^{l}_{\alpha}}
 \Biggl(
 \frac{q_{k,r}^{\alpha} - q_{l,r'}^{\alpha} - \ri\kappa}{
 q_{k,r}^{\alpha} - q_{l,r'}^{\alpha} }
 \Biggr)
 \prod_{\alpha<\beta}^{M} \prod_{k=1}^{N}\prod_{l=1}^{N}\prod_{r=1}^{m^{k}_{\alpha}} \prod_{r'=1}^{m^{l}_{\alpha}}
\Biggl(
 \frac{q_{k,r}^{\alpha} - q_{l,r'}^{\alpha} - \ri\kappa}{
 q_{k,r}^{\alpha} - q_{l,r'}^{\alpha} }
 \Biggr) \; .
\end{equation}

Substituting the expressions for $q_{k,r}^{\alpha}$, equation (\ref{23}), one can find an explicit
formula for the above factor $G_{M}$ which is rather cumbersome: it contains the products of all kinds
of the Gamma functions of the type
$\Gamma\big[1 +\frac{1}{2}\big(\sum_{k}^{N}(\pm)m^{k}_{\alpha} + \sum_{l}^{N}(\pm)m^{l}_{\beta}\big)
\pm \frac{1}{\kappa}(q_{\alpha}-q_{\beta})\big]$ [the example of this kind of the product is given in
\cite{end-point}, equation (A17)]. We do not reproduce it here as it turns out to be irrelevant in the limit
$t\to\infty$ (see below).

\begin{figure}[!b]
\centerline{
\includegraphics[width=0.4\textwidth]{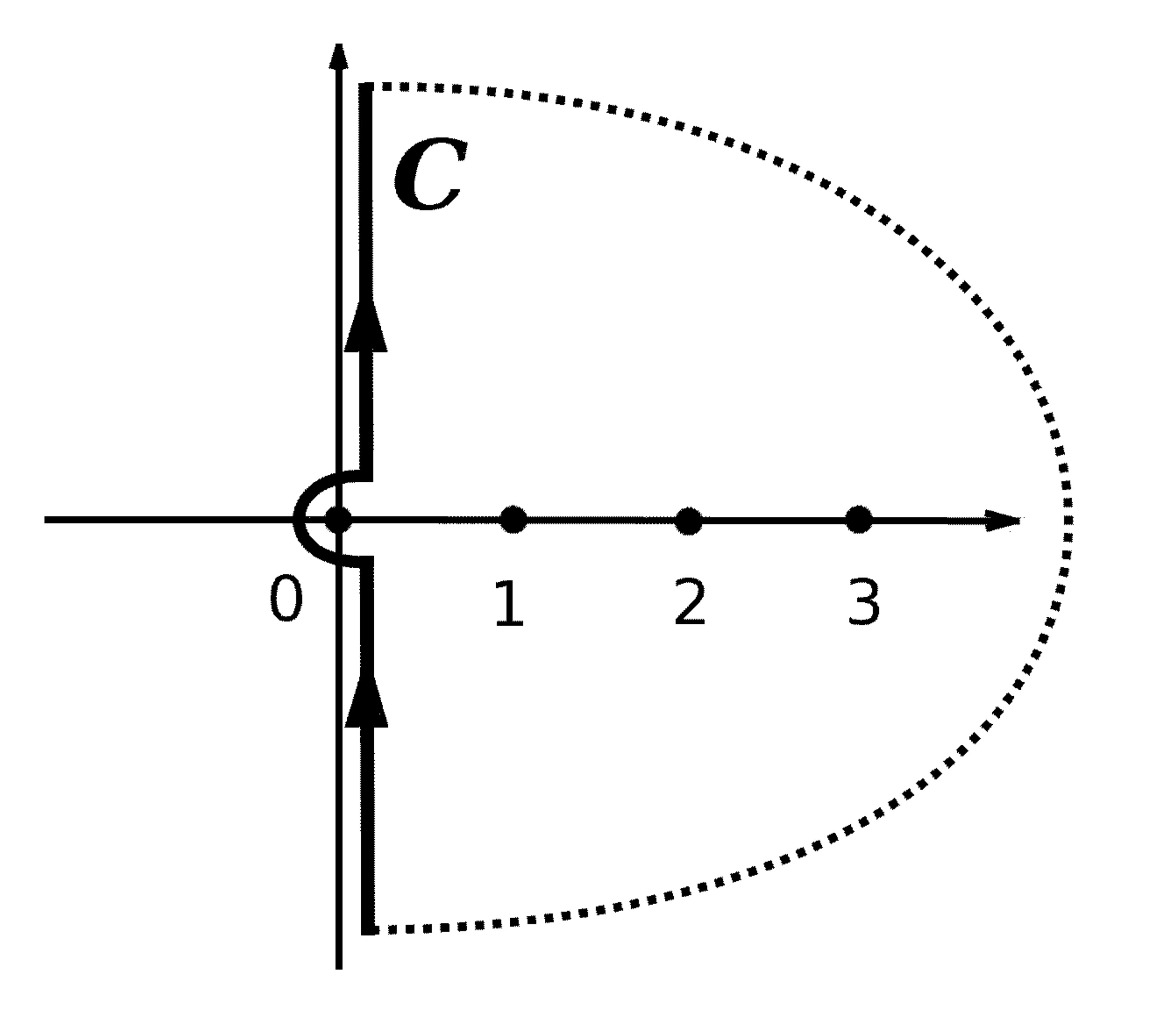}
}
\caption{The contours of integration in the complex plane used for
           summing the series equation (\ref{39}).\label{figure1}}
\end{figure}

After rescaling
\begin{eqnarray}
\label{26}
q_{\alpha} &\to& \frac{\kappa}{2\lambda} \, q_{\alpha}\,,
\\
\label{27}
x_{k} &\to&  \frac{2 \lambda^{2}}{\kappa} \, x_{k}\,,
\end{eqnarray}
with
\begin{equation}
\label{28}
\lambda \; = \; \frac{1}{2} \,
\left(\frac{\kappa^{2} t}{\beta}\right)^{1/3} \; = \;
\frac{1}{2} \, \left(\beta^{5} u^{2} t\right)^{1/3}
\end{equation}
the normalization factor $\big|C_{M}({\bf q}; \{m^{k}_{\alpha} \})\big|^{2}$, equation (\ref{16})
(with $n_{\alpha} = \sum_{k}^{N}m^{k}_{\alpha}$),
can be represented as follows:
\begin{eqnarray}
\nonumber
|C_{M}({\bf q}; \{m^{k}_{\alpha} \})|^{2} &=&
\prod_{\alpha<\beta}^{M}
\frac{
\big|
\lambda\sum_{k}^{N}m^{k}_{\alpha} - \lambda\sum_{k}^{N}m^{k}_{\beta} -
\ri q_{\alpha} + \ri q_{\beta}
\big|^{2}}{
\big|
\lambda\sum_{k}^{N}m^{k}_{\alpha} + \lambda\sum_{k}^{N}m^{k}_{\beta} -
\ri q_{\alpha} + \ri q_{\beta}
\big|^{2} }
\\
\nonumber
\\
\label{29}
&=&
\left[\prod_{\alpha=1}^{M}
\left(2\lambda \sum_{k}^{N}m^{k}_{\alpha}\right)\right]
\det
\Biggl[
\frac{1}{
\bigl(\sum_{k}^{N}\lambda m^{k}_{\alpha} - \ri q_{\alpha}\bigr) +
\bigl(\sum_{k}^{N}\lambda m^{k}_{\beta} + \ri q_{\beta}\bigr)}
\Biggr]_{\alpha,\beta=1,\ldots,M} \, .
\end{eqnarray}
Substituting equation (\ref{25})--(\ref{28}) into equation (\ref{23}) and using the Airy function relation
\begin{equation}
   \label{30}
\exp\left[ \frac{1}{3} \lambda^{3} \left(\sum_{k}^{N}m^{k}_{\alpha}\right)^{3} \right] \; = \;
\int_{-\infty}^{+\infty} \rd y \; \Ai(y) \;
\exp\left[\lambda \left(\sum_{k}^{N}m^{k}_{\alpha}\right) \, y \right]
\end{equation}
we get
\begin{eqnarray}
 \nonumber
W({\bf f}; {\bf x}) &=& 1 + \lim_{\lambda\to\infty} \left(
\sum_{M=1}^{\infty} \frac{(-1)^{M}}{M!}
\prod_{\alpha=1}^{M}
\left\{
\int\int_{-\infty}^{+\infty}\frac{\rd q_{\alpha}\rd y_{\alpha}}{2\pi} \Ai\left(y_{\alpha}+q_{\alpha}^{2}\right) \right.\right.
\\[1ex]
\nonumber
&&\times
\sum_{\sum_{k}^{N}m^{k}_{\alpha} \geqslant 1} (-1)^{\sum_{k}^{N}m^{k}_{\alpha} -1}
\exp
\left[
\lambda\sum_{k=1}^{N}m^{k}_
{\alpha} \left(y_{\alpha} + f_{k} +\ri x_{k} q_{\alpha} \right)
-\frac{1}{2} \lambda^{2} \sum_{k,l=1}^{N}m^{k}_{\alpha}m^{l}_{\alpha} \Delta_{kl}
\right]
\left.\vphantom{\int_{-\infty}^{+\infty}}
\right\}
\\[1ex]
&&\times
\det\hat{K} \left[
\left(\sum_{k}^{N}\lambda m^{k}_{\alpha}, \; q_{\alpha}\right) ; \; \left(\sum_{k}^{N}\lambda m^{k}_{\beta}, \; q_{\beta}\right)
\right]_{\alpha,\beta=1,\ldots,M}
G_{M}\left(\frac{\kappa {\bf q}}{2\lambda}; \; \{ m^{k}_{\alpha}\}\right)
\left.\vphantom{\int_{-\infty}^{+\infty}}
\right)\,,
\label{31}
\end{eqnarray}
where
\begin{equation}
 \label{32}
 \Delta_{kl} \; = \; \big|x_{k} - x_{l}\big|
\end{equation}
and
\begin{equation}
 \label{33}
 \hat{K} \left[
\left(\sum_{k}^{N}\lambda m^{k}_{\alpha}, \; q_{\alpha}\right) ; \; \left(\sum_{k}^{N}\lambda m^{k}_{\beta}, \; q_{\beta}\right)
\right] =
\frac{1}{\left(\sum_{k}^{N}\lambda m^{k}_{\alpha} - \ri q_{\alpha}\right) +
         \left(\sum_{k}^{N}\lambda m^{k}_{\beta} + \ri q_{\beta}\right)} \, .
\end{equation}
The quadratic in $m^{k}_{\alpha}$ term in the exponential of equation (\ref{31}) can be linearized as follows:
\begin{eqnarray}
 \nonumber
 \exp
 \left\{
 - \frac{1}{2} \lambda^{2}  \sum_{k,l=1}^{N}m^{k}_{\alpha}m^{l}_{\alpha} \Delta_{kl}
 \right\}
 &=&
 \exp
 \left\{
 - \frac{1}{4} \lambda^{2}  \sum_{k,l=1}^{N} \Delta_{kl} \left( m^{k}_{\alpha} + m^{l}_{\alpha}\right)^{2}
 + \frac{1}{2} \lambda^{2}  \sum_{k=1}^{N} \left(m^{k}_{\alpha}\right)^{2} \sum_{l=1}^{N} \Delta_{kl}
\right\}\nonumber
\\
\nonumber
&=&
\prod_{k,l=1}^{N}
\left\{
\int_{-\infty}^{+\infty} \frac{\rd \xi_{kl}^{\alpha}}{\sqrt{2\pi}}
\exp
 \left[
 -\frac{1}{2} \left(\xi_{kl}^{\alpha}\right)^{2}
 \right]
 \right\} \;
 \prod_{k=1}^{N}
\left\{
\int_{-\infty}^{+\infty} \frac{\rd \eta_{k}^{\alpha}}{\sqrt{2\pi}}
\exp
 \left[
 -\frac{1}{2} \left(\eta_{k}^{\alpha}\right)^{2}
 \right]
 \right\}
\\
&&\times
 \exp
 \left\{
\lambda  \sum_{k}^{N}\left[
\frac{\ri}{\sqrt{2}}\sum_{l=1}^{N} \sqrt{\Delta_{kl}} \, \left(\xi_{kl}^{\alpha} + \xi_{lk}^{\alpha}\right)
-\sqrt{\gamma_{k}} \, \eta_{k}^{\alpha}
\right] m^{k}_{\alpha}
\right\}\,,
\label{34}
\end{eqnarray}
where
\begin{equation}
 \label{35}
 \gamma_{k} \; = \; \sum_{l=1}^{N} \Delta_{kl} \; = \;
 \sum_{l=1}^{N}  \big|x_{k} - x_{l}\big| \; .
\end{equation}
Substituting the representation (\ref{34}) into equation (\ref{31}) and redefining the integration parameters
\begin{equation}
 \label{36}
 \eta_{k}^{\alpha} \; \to \;
 \eta_{k}^{\alpha} + \frac{\ri}{\sqrt{\gamma_{k}}} \, q_{\alpha} x_{k} +
 \ri \sum_{l=1}^{N} \sqrt{\frac{\Delta_{kl}}{2\gamma_{k}}} \; \bigl(\xi_{kl}^{\alpha} + \xi_{lk}^{\alpha}\bigr)
\end{equation}
we get
\begin{align}
 \nonumber
W({\bf f}; {\bf x}) =& 1 +
\sum_{M=1}^{\infty} \frac{(-1)^{M}}{M!}
\prod_{\alpha=1}^{M}
\left(
\int\int_{-\infty}^{+\infty}\frac{\rd q_{\alpha}\rd y_{\alpha}}{2\pi} \Ai(y_{\alpha}+q_{\alpha}^{2})
\prod_{k,l=1}^{N}
\left(
\int_{-\infty}^{+\infty} \frac{\rd \xi_{kl}^{\alpha}}{\sqrt{2\pi}}
\right)
\prod_{k=1}^{N}
\left(
\int_{-\infty}^{+\infty} \frac{\rd \eta_{k}^{\alpha}}{\sqrt{2\pi}}
\right)\right.
\\
&\times
\exp
 \left\{
 -\frac{1}{2} \sum_{k,l=1}^{N} \left(\xi_{kl}^{\alpha}\right)^{2}
 -\frac{1}{2} \sum_{k=1}^{N}
 \left[
 \eta_{k}^{\alpha} + \frac{\ri}{\sqrt{\gamma_{k}}} \, q_{\alpha} x_{k} +
 \ri \sum_{l=1}^{N} \sqrt{\frac{\Delta_{kl}}{2\gamma_{k}}} \; \left(\xi_{kl}^{\alpha} + \xi_{lk}^{\alpha}\right)
 \right]^{2}
 \right\}
 \left.\vphantom{\int_{-\infty}^{+\infty}}\right)
  {\cal S}\bigl({\bf f}, {\bf y}, {\bf q}, \{\eta_{k}\} \bigr)\, ,
 \label{37}
\end{align}
where
\begin{eqnarray}
 \nonumber
 {\cal S}\left({\bf f}, {\bf y}, {\bf q}, \{\eta_{k}\} \right) &=&
 \lim_{\lambda\to\infty}
 \prod_{\alpha=1}^{M}
 \left\{
 \sum_{\sum_{k}^{N}m^{k}_{\alpha} \geqslant 1} (-1)^{\sum_{k}^{N}m^{k}_{\alpha} -1}
\exp
\left[
\lambda \sum_{k=1}^{N} m^{k}_{\alpha} \left(y_{\alpha} + f_{k} - \sqrt{\gamma_{k}} \eta_{k} \right)
\right]\right.
\\
&&\times\left.
\det\hat{K} \left[
\left(\sum_{k}^{N}\lambda m^{k}_{\alpha}, \; q_{\alpha}\right) ; \; \left(\sum_{k}^{N}\lambda m^{k}_{\beta}, \; q_{\beta}\right)
\right]_{\alpha,\beta=1,\ldots,M}
G_{M}\left(\frac{\kappa {\bf q}}{2\lambda}; \; \{ m^{k}_{\alpha}\}\right)\right\} \, .
\label{38}
\end{eqnarray}
The summations over $m^{k}_{\alpha}$ in the above expression can be performed as follows:
\begin{align}
 \nonumber
 {\cal S}\left({\bf f}, {\bf y}, {\bf q}, \{\eta_{k}\} \right) =&
 \lim_{\lambda\to\infty}
 \prod_{\alpha=1}^{M}
 \left[
 \prod_{k=1}^{N}
 \left(
 \sum_{m^{k}_{\alpha}=0}^{\infty}  \delta_{m^{k}_{\alpha}, \, 0}
 \right)
 - (-1)^{N}
\prod_{k=1}^{N}
 \left\{
 \sum_{m^{k}_{\alpha}=0}^{\infty} (-1)^{m^{k}_{\alpha}-1}
 \exp
\left[
\lambda  m^{k}_{\alpha} \left(y_{\alpha} + f_{k} - \sqrt{\gamma_{k}} \eta_{k} \right)
\right]
\right\}
\right]
\\
\nonumber
&\times
\det\hat{K} \left[
\left(\sum_{k}^{N}\lambda m^{k}_{\alpha}, \; q_{\alpha}\right) ; \; \left(\sum_{k}^{N}\lambda m^{k}_{\beta}, \; q_{\beta}\right)
\right]_{\alpha,\beta=1,\ldots,M}
\times
G_{M}\left(\frac{\kappa {\bf q}}{2\lambda}; \; \{ m^{k}_{\alpha}\}\right)
\\
\nonumber
=&
\lim_{\lambda\to\infty}
 \prod_{\alpha=1}^{M}
 \left[
 \prod_{k=1}^{N}
 \left(
 \int_{{\cal C}} \rd z^{k}_{\alpha}  \, \delta(z^{k}_{\alpha})
 \right)
  - (-1)^{N}
\prod_{k=1}^{N}
 \left\{
 \int_{{\cal C}} \frac{\rd z^{k}_{\alpha}}{2\ri \sin(\pi z^{k}_{\alpha})}
 \exp
\left[
\lambda  z^{k}_{\alpha} \left(y_{\alpha} + f_{k} - \sqrt{\gamma_{k}} \eta_{k} \right)
\right]
\right\}
\right]
\\
&\times
\det\hat{K} \left[
\left(\sum_{k}^{N}\lambda z^{k}_{\alpha}, \; q_{\alpha}\right) ; \; \left(\sum_{k}^{N}\lambda z^{k}_{\beta}, \; q_{\beta}\right)
\right]_{\alpha,\beta=1,\ldots,M}
G_{M}\left(\frac{\kappa {\bf q}}{2\lambda}; \; \{ z^{k}_{\alpha}\}\right)\,,
\label{39}
\end{align}
where the integration goes over the contour ${\cal C}$ shown in figure~\ref{figure1}.
Redefining $z^{k}_{\alpha} \to z^{k}_{\alpha}/\lambda$, in the limit $\lambda\to\infty$, we get
\begin{eqnarray}
 \nonumber
 {\cal S}\left({\bf f}, {\bf y}, {\bf q}, \{\eta_{k}\} \right) &=&
 \prod_{\alpha=1}^{M}
 \left[
 \prod_{k=1}^{N}
 \left(
 \int_{{\cal C}} \rd z^{k}_{\alpha}  \, \delta(z^{k}_{\alpha})
 \right)
  - (-1)^{N}
\prod_{k=1}^{N}
 \left\{
 \int_{{\cal C}} \frac{\rd z^{k}_{\alpha}}{2\pi \ri \, z^{k}_{\alpha}}
 \exp
\left[
 z^{k}_{\alpha} \left(y_{\alpha} + f_{k} - \sqrt{\gamma_{k}} \eta_{k} \right)
\right]
\right\}
\right]
\\
&&\times
\det\hat{K} \left[
\left(\sum_{k}^{N} z^{k}_{\alpha}, \; q_{\alpha}\right) ; \; \left(\sum_{k}^{N} z^{k}_{\beta}, \; q_{\beta}\right)
\right]_{\alpha,\beta=1,\ldots,M}
\lim_{\lambda\to\infty}
G_{M}\left(\frac{\kappa {\bf q}}{2\lambda}; \; \{ \frac{z^{k}_{\alpha}}{\lambda}\}\right) \, .
\label{40}
\end{eqnarray}
Taking into account the Gamma function property $\lim_{|z|\to 0}\Gamma(1+z) = 1$,
one can easily demonstrate (see e.g. \cite{end-point})
that
\begin{equation}
 \label{41}
 \lim_{\lambda\to\infty}
G_{M}\left(\frac{\kappa {\bf q}}{2\lambda}; \; \left\{ \frac{z^{k}_{\alpha}}{\lambda}\right\}\right) \; = \; 1 \,    .
\end{equation}
Thus, in the limit $\lambda\to\infty$, the expression (\ref{37}) takes the form of the Fredholm determinant
\begin{eqnarray}
 \nonumber
W({\bf f}; {\bf x}) &=& 1 +
\sum_{M=1}^{\infty} \frac{(-1)^{M}}{M!}
\prod_{\alpha=1}^{M}
\left[
\int\int_{-\infty}^{+\infty}\frac{\rd q_{\alpha}\rd y_{\alpha}}{2\pi} \Ai(y_{\alpha}+q_{\alpha}^{2})
\prod_{k,l=1}^{N}
\left(
\int_{-\infty}^{+\infty} \frac{\rd \xi_{kl}^{\alpha}}{\sqrt{2\pi}}
\right)
\prod_{k=1}^{N}
\left(
\int_{-\infty}^{+\infty} \frac{\rd \eta_{k}^{\alpha}}{\sqrt{2\pi}}
\right)\right.
\\
\nonumber
&&\times
\exp
 \left\{
 -\frac{1}{2} \sum_{k,l=1}^{N} \xi_{kl}^{2}
 -\frac{1}{2} \sum_{k=1}^{N}
 \left[
 \eta_{k}^{\alpha} + \frac{\ri}{\sqrt{\gamma_{k}}} \, q_{\alpha} x_{k} +
 \ri \sum_{l=1}^{N} \sqrt{\frac{\Delta_{kl}}{2\gamma_{k}}} \; \left(\xi_{kl}^{\alpha} + \xi_{lk}^{\alpha}\right)
 \right]^{2}
 \right\}
 \\
\nonumber
&&\times\left.
 \prod_{k=1}^{N}
 \left(
 \int_{{\cal C}} \rd z^{k}_{\alpha}
 \right)
 \left\{
 \prod_{k=1}^{N} \, \delta(z^{k}_{\alpha}) \; - \; (-1)^{N}
 \prod_{k=1}^{N}
 \frac{1}{2\pi \ri z^{k}_{\alpha}}
 \exp
\left[
 z^{k}_{\alpha} \left(y_{\alpha} + f_{k} - \sqrt{\gamma_{k}} \eta_{k} \right)
\right]
 \right\}
 \right]
  \label{42}
\\
&&\times
\det\hat{K} \left[
\left(\sum_{k}^{N} z^{k}_{\alpha}, \; q_{\alpha}\right) ; \; \left(\sum_{k}^{N} z^{k}_{\beta}, \; q_{\beta}\right)
\right]_{\alpha,\beta=1,\ldots,M}
\\
&\equiv&
\det\left[\hat{1} - \hat{A}\right] \; = \; \exp\left\{- \sum_{M=1}^{\infty} \frac{1}{M} \textrm{Tr} \hat{A}^{M}\right\}\, ,
 \label{43}
\end{eqnarray}
where $\hat{A}$ is the integral operator with the kernel
\begin{eqnarray}
 \nonumber
 A\left[
 \left(\sum_{k}^{N} z^{k}, \, q\right); \; \left(\sum_{k}^{N} \tilde{z}^{k}, \, \tilde{q}\right)
 \right]
 &=&
 \int_{-\infty}^{+\infty}\frac{\rd y}{2\pi} \Ai(y+q^{2})
\prod_{k,l=1}^{N}
\left(
\int_{-\infty}^{+\infty} \frac{\rd \xi_{kl}}{\sqrt{2\pi}}
\right)
\prod_{k=1}^{N}
\left(
\int_{-\infty}^{+\infty} \frac{\rd \eta_{k}}{\sqrt{2\pi}}
\right)
\\
\nonumber
&&\times
\exp
 \left\{
 -\frac{1}{2} \sum_{k,l=1}^{N} \xi_{kl}^{2}
 -\frac{1}{2} \sum_{k=1}^{N}
 \left[
 \eta_{k} + \frac{\ri}{\sqrt{\gamma_{k}}} \, q_{\alpha} x_{k} +
 \ri \sum_{l=1}^{N} \sqrt{\frac{\Delta_{kl}}{2\gamma_{k}}} \; \left(\xi_{kl} + \xi_{lk}\right)
 \right]^{2}
 \right\}
 \\
\nonumber
&&\times
\left\{
 \prod_{k=1}^{N} \, \delta(z^{k}) \; - \; (-1)^{N}
 \prod_{k=1}^{N}
 \frac{1}{2\pi \ri z^{k}}
 \exp
\left[
 z^{k} \left(y + f_{k} - \sqrt{\gamma_{k}} \eta_{k} \right)
\right]
 \right\}
\\
&&\times
\frac{1}{\sum_{k}^{N} z^{k} \, - \, \ri q \; + \; \sum_{k}^{N} \tilde{z}^{k} \, + \, \ri \tilde{q}} \, .
\end{eqnarray}
Correspondingly, for the trace of this operator in the $M$-th power
[in the exponential representation of the
Fredholm determinant, equation (\ref{43})] we get
\begin{eqnarray}
 \nonumber
 \textrm{Tr} \hat{A}^{M} &=&
 \prod_{\alpha=1}^{M}
 \left[
 \int\int_{-\infty}^{+\infty}\frac{\rd y \rd q_{\alpha}}{2\pi} \Ai(y+q_{\alpha}^{2})
\prod_{k,l=1}^{N}
\left(
\int_{-\infty}^{+\infty} \frac{\rd \xi_{kl}}{\sqrt{2\pi}}
\right)
\prod_{k=1}^{N}
\left(
\int_{-\infty}^{+\infty} \frac{\rd \eta_{k}}{\sqrt{2\pi}}
\right)
\right.
\\
\nonumber
&&\times
\exp
 \left\{
 -\frac{1}{2} \sum_{k,l=1}^{N} \xi_{kl}^{2}
 -\frac{1}{2} \sum_{k=1}^{N}
 \left[
 \eta_{k} + \frac{i}{\sqrt{\gamma_{k}}} \, q_{\alpha} x_{k} +
 \ri \sum_{l=1}^{N} \sqrt{\frac{\Delta_{kl}}{2\gamma_{k}}} \; \left(\xi_{kl} + \xi_{lk}\right)
 \right]^{2}
 \right\}
\\
\nonumber
&&\times\left.
\prod_{k=1}^{N}\left(\int_{{\cal C}} \rd z^{k}_{\alpha} \right)
\left\{
 \prod_{k=1}^{N} \, \delta(z^{k}_{\alpha}) \; - \; (-1)^{N}
 \prod_{k=1}^{N}
 \frac{1}{2\pi \ri z^{k}_{\alpha}}
 \exp
\left[
 z^{k}_{\alpha} \left(y + f_{k} - \sqrt{\gamma_{k}} \eta_{k} \right)
\right]
 \right\}
 \right]
\\
&&\times
\prod_{\alpha=1}^{M}
\left(
\frac{1}{\sum_{k}^{N} z^{k}_{\alpha} \, - \, \ri q_{\alpha} \; + \;
         \sum_{k}^{N} z^{k}_{\alpha +1} \, + \, \ri q_{\alpha + 1}}
\right)\, ,
\label{45}
\end{eqnarray}
where, by definition, $z^{k}_{M+1} \equiv z^{k}_{1}$ and $q_{M+1} \equiv q_{1}$.

Substituting
\begin{equation}
 \label{46}
 \frac{1}{\sum_{k}^{N} z^{k}_{\alpha} \, - \, \ri q_{\alpha} \; + \;
         \sum_{k}^{N} z^{k}_{\alpha +1} \, + \, \ri q_{\alpha + 1}} \; = \;
\int_{0}^{\infty} \rd\omega_{\alpha}
\exp\left\{
-\omega_{\alpha} \left(
\sum_{k}^{N} z^{k}_{\alpha} \, - \, \ri q_{\alpha} \; + \;
\sum_{k}^{N} z^{k}_{\alpha +1} \, + \, \ri q_{\alpha + 1}
\right)
\right\}
\end{equation}
into equation (\ref{45}) we obtain
\begin{equation}
 \label{47}
 \mbox{Tr} \hat{A}^{M} \; = \;
 \int_{0}^{\infty} \ldots \int_{0}^{\infty} \rd\omega_{1} \ldots \rd\omega_{M} \;
 \prod_{\alpha=1}^{M}
 A\left(\omega_{\alpha} ; \; \omega_{\alpha + 1}\right)\,,
\end{equation}
where
\begin{eqnarray}
 \nonumber
 A\left(\omega ; \; \omega'\right)
 &=&
 \int\int_{-\infty}^{+\infty}\frac{\rd y \rd q}{2\pi} \Ai(y+q^{2} + \omega + \omega')
\prod_{k,l=1}^{N}
\left(
\int_{-\infty}^{+\infty} \frac{\rd \xi_{kl}}{\sqrt{2\pi}}
\right)
\prod_{k=1}^{N}
\left(
\int_{-\infty}^{+\infty} \frac{\rd \eta_{k}}{\sqrt{2\pi}}
\right)
\\
\nonumber
&&\times
\exp
 \left\{
 -\frac{1}{2} \sum_{k,l=1}^{N} \xi_{kl}^{2}
 -\frac{1}{2} \sum_{k=1}^{N}
 \left[
 \eta_{k} + \frac{\ri}{\sqrt{\gamma_{k}}} \, q x_{k} +
 \ri \sum_{l=1}^{N} \sqrt{\frac{\Delta_{kl}}{2\gamma_{k}}} \; \left(\xi_{kl} + \xi_{lk}\right)
 \right]^{2}
 -\ri q (\omega - \omega')
 \right\}
\\
&&\times
\left\{ 1 \;  - \; (-1)^{N}
 \prod_{k=1}^{N} \int_{{\cal C}}
 \frac{\rd z^{k}}{2\pi \ri z^{k}}
 \exp
\left[
 z^{k} \left(y + f_{k} - \sqrt{\gamma_{k}} \eta_{k} \right)
\right]
 \right\}\,.
 \label{48}
\end{eqnarray}
Integrating over $z^{1}, \ldots, z^{N}$, we finally get
\begin{eqnarray}
 \nonumber
 A\left(\omega ; \; \omega'\right)
 &=&
 \int\int_{-\infty}^{+\infty}\frac{\rd y \rd q}{2\pi} \Ai(y+q^{2} + \omega + \omega')
\prod_{k,l=1}^{N}
\left(
\int_{-\infty}^{+\infty} \frac{\rd \xi_{kl}}{\sqrt{2\pi}}
\right)
\prod_{k=1}^{N}
\left(
\int_{-\infty}^{+\infty} \frac{\rd \eta_{k}}{\sqrt{2\pi}}
\right)
\\
\nonumber
&&\times
\exp
 \left\{
 -\frac{1}{2} \sum_{k,l=1}^{N} \xi_{kl}^{2}
 -\frac{1}{2} \sum_{k=1}^{N}
 \left[
 \eta_{k} + \frac{\ri}{\sqrt{\gamma_{k}}} \, q x_{k} +
 \ri \sum_{l=1}^{N} \sqrt{\frac{\Delta_{kl}}{2\gamma_{k}}} \; \left(\xi_{kl} + \xi_{lk}\right)
 \right]^{2}
 -\ri q \left(\omega - \omega'\right)
 \right\}
\\
&&\times
\left[ 1 \;  - \; (-1)^{N}
 \prod_{k=1}^{N}
 \theta \left( -y - f_{k} + \eta_{k} \sqrt{\gamma_{k}} \; \right)
 \right]\,,
 \label{49}
\end{eqnarray}
where $\Delta_{kl} = \big|x_{k} - x_{l}\big|$ and $\gamma_{k} = \sum_{l=1}^{N}\Delta_{kl}$.

Thus, the $N$-point free energy distribution function
$W\bigl(f_{1}, \ldots, f_{N}; \; x_{1}, \ldots, x_{N}\bigr)$, equation (\ref{7}), is given by the Fredholm determinant
\begin{equation}
 \label{50}
 W\left({\bf f}; \; {\bf x}\right) \; = \; \det\left[ \hat{1} \; - \; \hat{A} \right]\,,
\end{equation}
where $\hat{A}$ is the integral operator with the kernel $A\left(\omega ; \; \omega'\right)$
(with $\omega, \omega' \geqslant 0$) represented in equation (\ref{49}).

\section{Conclusions}

In this paper using the method developed in \cite{2pointPDF} we extended our result to the spatial
$N$-point free energy distribution function in the thermodynamic limit $t\to\infty$.
It should be noted that following the ideas of the proof \cite{Imamura-Sasamoto-Spohn}
for the two-point function, one can easily demonstrate that the result (\ref{49})--(\ref{50})
obtained in this paper is equivalent to that derived earlier by Prolhac and Spohn \cite{Prolhac-Spohn-N}.
It should be stressed, however, that since
the obtained result for the kernel $A\left(\omega ; \; \omega'\right)$,
equation (\ref{49}), has a rather complicated structure, its analytic properties
are at present completely unclear and their study would require special efforts.

\vspace{-2mm}

\section*{Acknowledgements}

This work was supported in part by the grant IRSES DCPA PhysBio-269139.

\ukrainianpart

 \title{$N$-точкова функція розподілу вільної енергії \\ в одновимірних хаотично напрямлених полімерах}

 \author[]{В. Доценко\refaddr{label1,label2}}
 \addresses{
 \addr{label1} Університет м. Париж VI, 75252 Париж, Франція \
 \addr{label2} Інститут теоретичної фізики ім. Л.Д. Ландау, 119334 Москва, РФ
 }

 \makeukrtitle

 \begin{abstract}
 \tolerance=3000%
 Отримано явний вираз для $N$-точкової функції розподілу вільної енергії в одновимірному напрямленому полімері в
 термінах анзацу Бете в рамках методу реплік. Отриманий результат еквівалентний результату, раніше отриманому в роботі
 Пролака і Шпона [J. Stat. Mech., 2011, P03020].
\keywords напрямлені полімери, хаотичний потенціал, репліки, флуктуації, функція розподілу

 \end{abstract}

\end{document}